\documentclass[10pt]{article}
\usepackage[utf8]{inputenc}
\usepackage{amsmath}
\usepackage{amssymb}
\usepackage{graphicx}
\usepackage{hyperref}
\usepackage[style=phys]{biblatex}
\begin{document}

\title{Probabilistic Reconstruction of Paleodemographic Signals}

\author{L.M. Arthur$^1$\footnote{\href{mailto:lmarthur@mit.edu}{lmarthur@mit.edu}} \and F. Chelazzi$^2$ \and D. Lawrence$^3$ \and M.D. Price$^3$}
\date{
    \small{
    $^1$Department of Nuclear Science and Engineering\\Massachusetts Institute of Technology\\[2ex]
    $^2$Department of Archaeology\\University of Glasgow\\[2ex]
    $^3$Department of Archaeology\\Durham University\\[2ex]
    } May 31, 2024
}

\maketitle

\abstract{We present a comprehensive Bayesian approach to paleodemography, emphasizing the proper handling of uncertainties. We then apply that framework to survey data from Cyprus, and quantify the uncertainties in the paleodemographic estimates to demonstrate the applicability of the Bayesian approach and to show the large uncertainties present in current paleodemographic models and data. We also discuss methods to reduce the uncertainties and improve the efficacy of paleodemographic models.}

\tableofcontents

\section{Introduction}
The fundamental obstacle to paleodemography is that it is not possible to directly observe populations of interest. To answer demographic questions, information must be extracted from indirect proxies. These proxies are often subject to large sampling and loss-rate uncertainties, and difficult-to-quantify correlations with confounding variables.\cite{chamberlain_demography_2006} The process of paleodemography then becomes one of extracting information and estimating parameters from noisy, incomplete, and confounded data, and of estimating the uncertainties in the inferred parameters. 

A wide set of proxies have been proposed, and used in different regions and periods with varying degrees of success. Among the most widely applicable are settlement data and summed probability distributions of radiocarbon dates. Settlement data can be structured in a variety of ways, including the number of settlements, the total settled area, or the number of residential structures within a region. Converting raw counts into time series poses difficulty due to sampling biases, preservation, and problems of contemporaneity. Traditional approaches to paleodemography from settlement data do not attempt to control for preservation, often neglect sampling effects, and do not rigorously account for or estimate uncertainties.\cite{drennan_demography_2015}

Over the past decade, summed probability distributions of radiocarbon dates have begun to be used as an additional proxy for population. The assumption here is that the number of dated events is a measure of human activity at any given point, and that human activity is correlated with population. However, radiocarbon dates are subject to many of the same biases and uncertainties as settlement data, with additional complications due to non-random collection of $^{14}$C samples in the course of excavation and submission of samples to radiocarbon laboratories. The number of radiocarbon dates obtained at a particular site depends as often on the period of occupation, the excavator's research questions, and the excavation budget. While these problems have been identified in the literature, they have not been addressed in a systematic way, and the uncertainties in the resulting population estimates are often not quantified.\cite{crema_inference_2021} Approaches such as binning, in which samples are grouped, often by site or time period, and thinning, in which a reduced set of samples are taken from a given site or bin, can be used to reduce the effects of sampling biases. But, these methods can introduce their own biases and uncertainties and necessarily discard information that could otherwise be used to constrain the population estimates. Other techniques such as logistic and Gaussian process regression can be used to smooth out noisy data, but this again reduces model expressivity and can introduce biases.

However, parameter and uncertainty estimation is well-studied in statistics and the natural sciences, and Bayesian methods are well-suited to the problem. Unlike other approaches to data analysis, the Bayesian approach allows the incorporation of all available information into a single model, and does not require the discarding of data in producing probability distributions for the inferred parameters. The resulting distributions mandate the propagation of uncertainties from the data to the final result, and the Bayesian approach allows for a clear accounting of the assumptions made in each step of data collection and analysis. This allows the inclusion of suspect, anomalous, or incomplete data, which maximizes the amount that can be learned from the data and limits the effect of outliers or of bias induced by the manual selection of data based on artificial criteria.\footnote{For an introductory discussion of Bayesian methods, see\cite{jaynes_probability_2003}. For a more comprehensive discussion, see \cite{gelman_bayesian_2020}.}

In the Bayesian treatment, uncertainty corresponds with the delocalization of probability density. The more uncertain a parameter, the more spread out its probability distribution. These distributions can correspond to physical properties inherent to the system, for example the number of people living within a given area at a given time, or they can correspond to uncertainties in nonphysical model parameters, such as the loss rate or the probability of a site being discovered or excavated.

Error and uncertainty come from two primary sources: The first is from inherently stochastic behavior, the effect of which can be reduced by increasing the quantity of data. The second is from assumptions about the methodology or about the system: for example, the assumption that the intensity of archaeological surveying is constant over space, or that human behavior and the attenuation of the archaeological record are constant over time and space. These assumptions are known to be incorrect, but are necessary to expedite the analysis and to make the problem tractable. Making an assumption introduces uncertainty that cannot be easily quantified, however, treating assumptions as parameters in a Bayesian framework, combined with the use of multiple data sources, allows for the quantification of these uncertainties and forces the mutual calibration of different sources of data. The Bayesian approach forces us to be explicit about assumptions which are often ignored in archaeological analyses.

With paleodemography there are behavioral parameters such as the likelihood that a given number of people in a time and place will leave behind a given quantity of some proxy that can be measured. For example, the number of people per habitation may change over time. If a relevant demographic proxy is the number of habitations, then the number of people per habitation is a crucial parameter in estimating the population size. Similarly, the intensity of the use of cooking fires, the quantity of pottery produced, the number of burials, and the quantity of material in midden deposits are all behavioral parameters that cannot be assumed to be constant.

There are also loss effects, which dictate the likelihood that a given proxy that has been deposited will remain intact for later discovery and study. This depends on the local environment. Relevant factors include rates of erosion and sedimentation, soil chemistry, the presence of scavengers, the disturbance of the site by subsequent human and animal activity, and other environmental changes such as sea level fluctuations, all of which vary over time and space.

Methodological uncertainties must also be addressed. These include sampling effects, such as the probability of a site containing relevant proxies being discovered, the probability of that site being excavated, and, in the case of radiocarbon dates, the probability of a sample being dated. These probabilities are not constant over time and space and may be difficult to quantify. For example, the probability of site discovery may depend on the intensity of archaeological surveys, ground vegetation, terrain, and land use. 

These uncertainties are correlated with each other, and cannot be averaged away, meaning that increasing the quantity of data without controlling for the individual uncertainties will not yield accurate results. However, they can be treated rigorously with Bayesian probabilistic methods.

\section{The Paleodemographic Paradigm}
In the Bayesian approach, a distribution $P(A|B)$ represents knowledge about a proposition $A$, given some condition or set of conditions $B$. The formalism is based on the consequences of two axioms---the sum rule and product rule---which form the foundation of probability theory and lead directly to Bayes' theorem and Bayesian methods.
\begin{equation}
    p(A|B, C) = \frac{p(A, B|C)}{p(B|C)} = \frac{p(B|A, C)p(A|C)}{p(B|C)}
\end{equation}

To apply Bayesian methods to complex systems, and to examine the implications of assumptions in the analysis, it is often necessary to construct a hierarchical model. This is a model in which some or all of the model parameters themselves are treated as stochastic variables, with beliefs about the values of these parameters described by probability distributions. Calculating posterior distributions from arbitrary prior distributions is not in general possible analytically, but can be done using numerical methods such as Markov chain Monte Carlo (MCMC) or variational inference (VI).

To structure the paleodemographic problem in a tractable form, we begin in the continuous approximation, treating population density as a continuously differentiable field $\psi(x, t)$, varying in space $x$ and time $t$. To extract a population estimate for a given time in a given region $A$, we integrate the population density over the relevant area.\footnote{Subscripts denote discrete variables, such as a particular region or time period, and parentheses denote continuous variables. The discretized variables are analogous to the data analysis concept of binning.}
\begin{equation}
    N_A(t) = \int_{A} \psi(x, t) dx
\end{equation}

Because the population density is not directly observable, we concern ourselves with proxies, which are observable quantities that are related to the population density. We denote a proxy field as $\phi^i(x, t)$, where the superscript index $i$ denotes a particular proxy. It is necessary to distinguish between related concepts: the rate of proxy deposition at a particular time and place, and the amount of proxy material existing at a particular time and place, dating from some other,earlier time. The local rate of deposition is a function of the local population density, with a form that may or may not be known. For example, we may assume that the rate of burials scales linearly with population density, while the number of cooking fires, or the amount of pottery produced, may scale nonlinearly with the population density by some scaling function $f^i$. We also assume locality---that the rate of proxy deposition at a particular time and place is only dependent on the population density at that time and place, and not on the population densities at other times or other places.
\begin{equation}
    \phi^i(x, t) = f^i(\psi(x, t))
\end{equation}
The function $f^i$ is assumed to follow a power law with real constants $a^i$ and $b^i$.

\begin{equation}
    f^i(\psi(x, t)) = a^i \psi(x, t)^{b^i}
\end{equation}

Because the time differences between when each of the excavations and surveys were carried out are small compared to the elapsed times between the proxy depositions and observations, we treat the observations of the proxy field as all being contemporaneous, occurring at the observation time $t_0$. We denote the field of proxies deposited at time $t$ and remaining at time $t_0$ as $\phi_0^i(x, t)$.

The relationship between $\phi_0^i(x, t)$ and $\phi^i(x, t)$, the initially deposited proxies from a given time is based on a process of loss over time. The magnitude of the total loss is a function of the time difference between the time of deposition and the observation time, and may vary widely between different proxies and between different regions. Relevant factors include erosion, weathering, sea level rise, sedimentation, human activity, soil chemistry, climate, bioturbation, and many others. We model the loss as an exponential decay process, with a decay rate $\lambda^i(x)$ that varies as a function of space. 
\begin{equation}
    \phi_0^i(x, t) = \phi^i(x, t) e^{-\lambda^i(x) \Delta t}
\end{equation}
The elapsed time between deposition and observation is given by $\Delta t = |t_0 - t|$. The number of proxies deposited at time $t$, which can in principle be observed in a given area at time $t_0$ is then given by integrating over the relevant region $A$.
\begin{equation}
    \Phi_{0, A}^i(t) = \int_A \phi_0^i(x, t) dx = \int_A \phi^i(x, t) e^{-\lambda^i(x) \Delta t} dx
\end{equation}

While $\Phi_{0, A}^i(t)$ is in principle an observable quantity, actual observations are further convolved with sampling distributions $p^i(\text{s} | x, t)$. This includes the probability that a particular site is discovered, and the probability that the site will be excavated, allowing the proxies to be observed. For some proxies there may be additional convolutions, such as for radiocarbon dates, where the probability that a given sample is sent to the lab and dated successfully. The data that is observed in a given area $A$, from a given time period $T$, is given by the complete convolution integral.
\begin{equation}
    \begin{split}
    \Phi_{T, A}^i =& \int_T \int_A \Phi_{0, A}^i(t) p^i(\text{s} | x, t) dx dt
    \\ =& \int_T \int_A a^i \psi(x, t)^{b^i} e^{-\lambda^i(x) \Delta t} p^i(\text{s} | x, t) dx dt
    \end{split}
\end{equation}
From the observed data $\Phi_{T, A}^i$, we must then perform the accompanying deconvolution, based on the parameters of the model, to obtain the population density $\psi(x, t)$.

\section{Uncertainty}
A key advantage of the Bayesian approach to data analysis is that it provides a natural framework for the quantification of uncertainty. Instead of reporting a single best estimate of the parameter of interest, and then facing the difficult task of quantifying the error in that estimate, the Bayesian approach provides a probability distribution for the parameter of interest, which encodes all of the relevant information and can be used to generate any desired summary statistics.

When testing predictions made by a Bayesian model, it is important to distinguish between errors that are due to the uncertainties in the model parameters, and errors that are due to the structure of the model itself. For example, if the paleodemographic model assumes locality of proxy deposition, but true deposition is non-local, or if the exponential decay model of loss is not a good approximation, then the model is not likely to fit the data. This is not a failure of the method, but an important feature. This allows the testing of the assumptions that went into the model. The discrepancy between the model and the data provides insight into the validity of the assumptions, and the comparison of different models---for example one that assumes exponential loss and one that assumes hyperbolic loss---can be used to determine which model most closely corresponds with reality.

Model predictions may be highly sensitive to certain parameters or structural assumptions, and less sensitive to others. Understanding this sensitivity is necessary for evaluating the validity of the assumptions and the predictions. For example, uncertainty in the loss rate $\lambda$ has a large contribution to the final uncertainty in the result due to the exponential nature of the loss process, and predictions may be more sensitive to the exponent $b$ than to the pre-factor $a$ in the assumed power law relationship between the population density and the proxy density. If the model is highly sensitive to a particular parameter, then constraining that parameter with data will significantly reduce the uncertainty in the final prediction. Consideration of parameter sensitivities can then guide the collection and analysis of new data.

\section{Case Study: Cyprus}
\label{sec:cyprus}
To demonstrate the applicability of the framework, and to emphasize the uncertainties that dominate paleodemographic estimates, we apply the probabilistic method to a case study from a single island, Cyprus, with no spatial considerations incorporated into the model. The Cyprus Settlement Dataset is available from Crawford and Vella under a CC-BY 4.0 license.\cite{crawford_cyprus_2022} The dataset comprises 1559 settlements on the island of Cyprus, spanning from the Late Epipaleolithic (11,000 BCE) to the end of the Ottoman period (1878 CE), collected from large-scale surveys and grey literature.\cite{catling1963, stanleyprice1980} We preprocessed this data into a time series of the number of occupied settlements, to be treated as a proxy for population, which is integrated over the entire island based on the assumption that all model parameters are constant over the geographic area of the island. We then restricted the time domain to end at 1000 CE, as many Byzantine, Frankish, Venetian, and Ottoman sites have remained occupied into the present day, resulting in their exclusion from the survey data.

Cyprus has functioned as an important gateway connecting the Mediterranean world with the major continental regions of Turkey and Southwest Asia. The island has held cultural and economic significance, both in antiquity and contemporary times. Cypriot society has existed in a challenging landscape with limited natural resources and a fragile ecosystem shaped by precipitation. While the impacts of climate change on Cyprus bear resemblance to those documented across Southwest Asia, localized responses to climatic disturbances have spawned unique cultural trajectories. Characterized by a deliberate cultivation of insular identity since its nascent stages, Cypriot culture stands out for its amalgamation of vulnerability, resilience, and capacity for transformational adaptation.\cite{clarke2003}

The Bayesian analysis was performed using the Python language and the Pyro probabilistic programming library.\cite{bingham2018pyro} The Pyro library provides a flexible framework for specifying probabilistic models, and for performing inference using both variational inference and Markov chain Monte Carlo (MCMC) methods. We implemented a stochastic variational inference (SVI) approach, which recasts the inference problem as an optimization problem, and used the gradient-based Adam optimizer to find the optimal model parameters. Because no relation between time steps $t$ and $t+1$ is assumed a priori, the dimensionality of the problem scales directly with the number of time steps in the data. We implemented SVI for this analysis, as it is more efficient than MCMC in this high-dimensional context.

\subsection{Model Construction}
We modeled the distribution for the population at each time step as a set of gamma distributions, with the initial modes for each distribution parameterized by an exponential function in time. We estimated the initial population at 11,000 BCE as 1000, and set the exponential growth rate to correspond with the 1881 census-reported population of 186,173.\cite{cystat1881} The initial standard deviations of the prior distributions were taken to be equal to the modes, to reflect large initial uncertainties. 

We also treated the loss rate as gamma distributed, with an arbitrarily chosen initial mode and standard deviation of 0.0001, corresponding to a constant loss of 10\% of settlements every 1000 years. Order-of-magnitude sensitivity checks indicated that the model was not sensitive to the initial choice of the loss rate.

For the scaling law between the population and the number of settlements, we restricted the exponent to unity, assuming a linear relationship. We modeled the pre-factor prior as a gamma distribution with an initial mode of 150 people, corresponding to Dunbar's number, and an often-assumed population for small settlements in the region.\cite{DUNBAR1992469,Wilkinson2010} The initial standard deviation was taken to be 150 as well, which corresponds to an expectation of approximately 60 settlements of 150 people for every settlement of 10,000 people.

The sampling probability for the settlements in the region is a function of the survey efficacies and the relationship between the survey areas and the total area under consideration. It is additionally complicated due to selection effects, where regions expected to have the highest densities of settlements are preferentially surveyed, meaning that the survey area is not a random sample of the landscape. For the combined Cyprus survey data, we assumed a beta distributed prior for the composite sampling probability, with a mode and standard deviation of 0.1.

\subsection{Model Results}
\begin{figure}[ht]
    \centering
    \includegraphics[width=\textwidth]{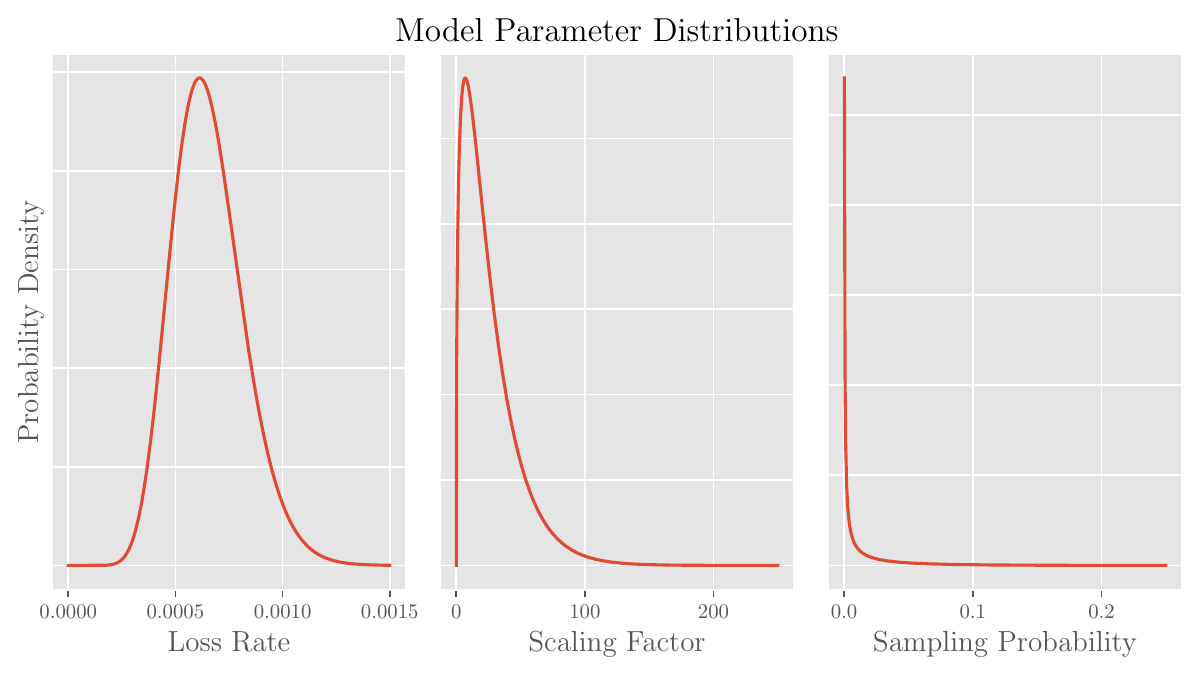}
    \caption{The posterior distributions for the loss rate $\lambda$, the scaling factor $a$, and the sampling probability $p$ for the Cyprus case study
    represent the combined information from the prior distributions and the survey data.}
    \label{fig:cyprus_posteriors}
\end{figure}

We generated a set of posterior distributions for the estimated populations at each time step and all model parameters via stochastic variational inference for 25,000 iterations of the Adam optimizer with a learning rate of 0.001. The posteriors for the model parameters are plotted in Figure \ref{fig:cyprus_posteriors}, and parameter summary statistics are provided in Table \ref{tab:cyprus_posteriors}. 

\begin{table}[ht]
    \centering
    \begin{tabular}{|l|rr|}
        \hline
        Parameter & Mean & Std. Dev. \\
        \hline
        Loss rate ($\lambda$) & 0.00065 & 0.00017 \\
        Scaling factor ($a$) & 25.78 & 22.03 \\
        Sampling probability ($p$) & 0.01 & 0.03 \\
        \hline
    \end{tabular}
    \caption{The summary statistics for the posterior distributions provide numerical estimates for the model parameters.}
    \label{tab:cyprus_posteriors}
\end{table}

\begin{figure}[ht]
    \centering
    \includegraphics[width=\textwidth]{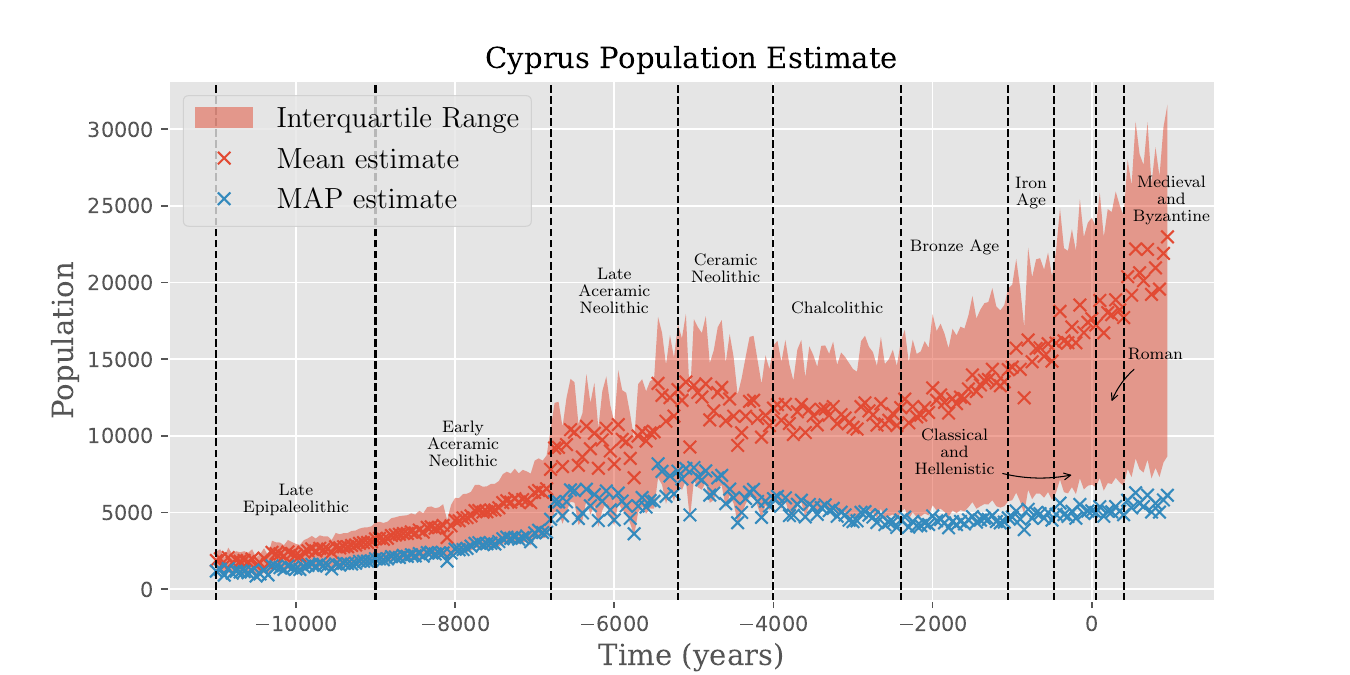}
    \caption{Large uncertainties in the population estimates are visible, however the general trend of increasing population 
    over time is clear in both the mean and maximum \textit{a posteriori} (MAP) estimates.}
    \label{fig:cyprus_population}
\end{figure}

The mean and maximum \textit{a posteriori} (MAP) estimates for the population at each time step are plotted in Figure \ref{fig:cyprus_population}. The shaded region represents the interquartile range of the posterior distributions, corresponding to 50\% probability that the true population at each time step lies within the shaded region. The uncertainties in the population estimates are large, however the general trend of increasing population over time is clear, as is a probable period of more rapid population growth around 7000 BCE, with a period of possible decline or stagnation between 5000 BCE and 2000 BCE, followed by a period of more rapid growth potentially corresponding to the establishment of complex societies on the island.

Because of the size of the uncertainty, the effects of short-term demographic changes caused by events such as natural disasters, epidemics, or warfare, are difficult to detect without model refinement and the addition of more data. However, the general trends in the population over time are clear, and the case study illustrates the utility of uncertainty estimates in avoiding the misinterpretation of noise in the data as signal.

\subsection{Model Discussion}
Our model indicates that the Early Holocene in Cyprus opens with a period of stable but moderate population growth, which aligns well with the commonly accepted hypothesis that the island witnessed several colonizing events when communities of farmers-herders arrived on its shores from the coastal regions of the Levant and Anatolia.\cite{simmons2004bitter,simmons2011rewriting, vigne2012first,Bradshaw2024}

Around 7000 BCE (Late Aceramic Neolithic), the demographic trend indicates a growth phase that might have resulted from the complete adoption of agropastoralism on the island. Archaeobotanical and zooarchaeological data suggest that during this period, the pattern in plant and animal data diverges from the trajectory observed on the mainland Levant, shifting towards a more diverse but grain-poor subsistence economy founded on small-scale intensive garden cultivation combined with livestock herding, capable of supporting a growing population.\cite{bolger2013matter,bogaard2005garden}

Demographic responses to the so-called 8.2k yr BP event vary across Southwest Asia, with some areas experiencing population growth and some stability or decline, perhaps as a result of movements into wetter regions.\cite{palmisano2021holocene} On Cyprus, repeated sub-centennial fluctuations in rainfall between moist and dry conditions between 7500 and 6000 BCE are visible in high-resolution climatic records from Jeita and Soreq caves.\cite{cheng2015climate,bar2003sea} This may have impeded the demographic growth we might expect after the onset of sedentism and the adoption of the full Neolithic package. While Cyprus may not strictly conform to the conventional definition of marginal environments (i.e. those incapable of sustaining uninterrupted rain-fed agriculture due to annual rainfall levels below the 250 mm/yr threshold), it effectively assumes characteristics akin to such environments during periods of drought. This is because successive years of diminished rainfall disproportionately impair agricultural yields and vegetative growth. Cyprus lacks accessible deep aquifers and significant standing bodies of water, and the rain-fed rivers stemming from the Troodos mountains are deeply incised, leading to rapid and forceful dispersion of water.\cite{clarke2016climatic,christodoulou1959evolution}

The 'long' 4th millennium BC (from the mid-5th millennium to the end of the 4th millennium BCE), characterized by the frequent occurrence of episodes of relative aridity, appears to coincide with a period of general demographic stagnation. The abandonment of villages at the beginning of the Chalcolithic period seems to align with a period of particular aridity, and a transition from agro-pastoralism to deer hunting, perhaps as a response to resource depletion and stress induced by the changing climate.\cite{clarke2016climatic} While some have suggested a reduction in scale or reliance on agriculture, other scholars have noted indications of agricultural diversification, potentially resulting from changing climatic conditions and demonstrating proficiency in sustaining reduced population densities.\cite{lucas2020against} This pattern underwent a dramatic shift at the onset of the Late Chalcolithic period (first half of the 3rd millennium BCE), as hunting began to decline, and the cultivation of cash crops and caprine husbandry began to rise. This new subsistence pattern would go on to characterize the entire subsequent Bronze Age.\cite{lucas2020against} Model results indicate a paleodemographic shift around the beginning of the 2nd millennium BCE, which coincides with a period of central importance in Cypriot prehistory marked by economic innovations and transformations of socio-economic networks, including the reintroduction of cattle, the introduction of equids and the plough, and the adoption of new copper technologies.\cite{webb2007identifying} The origin of these innovations on the island has been subject to various interpretations by archaeologists, ranging from the arrival of Anatolian refugees, more general patterns of ethnic migration and population movement in the Eastern Mediterranean, or hybridizing practices within a larger regional network.\cite{catling1971cyprus, webb2007identifying, webb2011hearth, knapp2013archaeology}

The reintroduction of cattle and improvements in metal object production suggest the expansion of agricultural fields into new areas with lower rainfall and poorer soils, yet in closer proximity to the island's copper resources. The entire Bronze Age is characterized by an expanding population and a shift from small-scale intensive to large-scale extensive cultivation systems. Evidence of this includes the impressive quantities of large storage vessels found in the main administrative centres of the Late Bronze Age, indicating complex staple finance systems characterized by practices of collection and redistribution of staple commodities, and complex settlement patterns where large sites were supported by a growing population settled in smaller industrial and rural sites.\cite{keswani1993models,andreou2016understanding}

A similar economic organization persisted into the subsequent Iron Age, marked by rural expansion and the emergence of urban centers that served as focal points for the main Cypriot city-kingdoms. It is noteworthy that our demographic model does not provide evidence for possible population decline resulting from the interplay of environmental, political, and economic events at the end of the Late Bronze Age across the Mediterranean region.\cite{kaniewski2019drought, knapp2016crisis} Our model prior reflects the 1881 census figure, an assumed low population in the Late Epipaleolithic, and approximately exponential growth over the intervening period. The relative sparsity of data from the Bronze Age onward in the Crawford and Vella dataset means that there is little additional information from the dataset to convey short timescale population dynamics to the model, and no such dynamics are reflected in the population estimate. Instead, during periods with scarce data, the model relies on the priors to produce a population estimate, and indicates larger uncertainties due to the lack of constraining data.

Although the city-kingdoms of Cyprus developed through different geo-linguistic and political trajectories from the Cypro-Geometric to the Classical periods, the Cypriot landscape became densely populated with multiple communities under the control of central authorities: dispersed farmsteads in the alluvial lowlands, productive sites near mineral and timber resources, and rural sanctuaries for territorial claims.\cite{iacovou2013historically, iacovou2018from, papantoniou2013cyprus}

During the Roman period, the interquartile range assumed by our model spans from ~7500 to ~25000, likely not capturing significant variations over short timescales. Archaeological data and textual sources indicate a complex population history for Cyprus in this period, with local dynamics intersecting with major changes in the broader geopolitical landscape of the Roman world.\cite{gordon2018transforming,rautman2000busy} Although characterized by cycles of population decrease (e.g., 3rd century CE) and increase (e.g., 6th to mid-7th centuries CE), Cyprus appears to have been a politically stable and integrated province, with urban activity documented in all major cities and expanding rural lifeways.\cite{vionis2022boom,rautman2000busy} It was only with Arab raids in the mid-7th to 9th centuries CE that long-established exchange routes were disrupted and political boundaries redrawn, while the urban and rural life of Cyprus was not significantly dislocated.\cite{zavagno2012edge}

It is further worth noting that these short timescale variations are not reflected in traditional summed probability distribution methods. Because radiocarbon dating is less common for sites with well-established ceramic typologies, non-Bayesian approaches that rely on radiocarbon dates as a population proxy are likely to underestimate the population for these sites and periods. Summed probability distributions then produce incorrect population estimates with no quantification of the uncertainties. While the Bayesian approach also produces estimates that are likely to be incorrect for periods with scarce data, the large uncertainties are quantified and clearly indicate that short timescale fluctuations may be present, but that additional data would be required to resolve any additional dynamics. 

\section{Conclusion}
The framework presented here provides a general method for estimating sizes and distributions of populations in the past. The framework is flexible and allows for extensions beyond the simple model demonstrated in the case study to capture more complex phenomena by increasing model expressivity, and to reduce uncertainties by incorporating additional data and prior information.

A clear approach to reducing uncertainties in the model and improving the sensitivity to demographic changes is to increase the quantity and quality of the data. This can be done by increasing survey coverage, such as through the use of extensive survey techniques reliant upon remote sensing, or by integrating more individual surveys. Additional intensive surveys can also increase the quantity of data and can be used to calibrate the sampling probabilities for the extensive surveys and to better estimate the distribution of settlement sizes in a region as a function of time.\cite{Knodell2022} Another key element of data quality is ensuring the accuracy and precision of site chronology. Aspects of this include increasing the quantity of radiocarbon dates for sites and calibrating stratigraphy and artifact typologies to increase the quantity of available chronological data and better constrain and account for chronological uncertainty.

Because the loss rates, scaling factors, and scaling exponents will differ between proxies, these contributions to final estimate uncertainty can be reduced by increasing the variety of different proxies identified in a region. For example, paleodemographic estimates from settlement data and from radiocarbon summed probability distributions have been generated independently and then compared, but they could instead be integrated into the same probabilistic model to extract information from both proxies.\cite{Palmisano2017,palmisano2021holocene} All else being equal, the contribution from each of these sources to the final estimate uncertainty will scale inversely with the square root of the number of proxies under consideration. 

Estimates can also be improved and uncertainties can be reduced through better model construction. For example, better priors for model parameters can be determined either through reducing the uncertainties in individual model parameter distributions or through identifying functional forms for the distributions that better match prior knowledge. Increasing the depth of the hierarchical elements of the model can also allow for more precise control over confounding variables, although increasing model dimensionality also increases the quantity of data required for meaningful inference. Using non-parametric density estimation techniques such as MCMC also yields an advantage by increasing the expressivity of the posterior distributions, not limiting them to a particular family of distributions as is the case in variational approaches. 

The sampling probabilities can be better constrained through increasing the fraction of the area under consideration that has been surveyed for the relevant proxy. Other calibration techniques can also be used to estimate sampling probabilities for specific surveys, such as through the comparison of different techniques in surveys with overlapping areas. Surveys and excavations from a variety of landscapes and environments can be used to attempt to control for biases in the geographic distribution and for variations in land-use, landscape productivity, and terrain.

Incorporating more specific prior knowledge about the behavior of populations through the development of parametric population models holds particular promise for more efficient and accurate inference. For example, the simple model discussed in Section \ref{sec:cyprus} assumes no relationship between the population at time $t$ and the population at time $t+1$. Incorporating a parametric model without artificially limiting expressivity, as is the case in standard logarithmic or exponential regression, poses a challenging task.

In addition to enforcing a relationship between time $t$ and time $t+1$, a natural extension of the framework would be to incorporate a relationship between position $x$ and position $x+1$. This would allow for treatment of the locality assumption and for the capture of shifting population centers, for example due to local environmental variability. This could be done through the subdivision of the region under consideration into some grid or other tiling and the imposition of a distance metric. The model could then capture not only increases and decreases in population at one particular point or region, but movement of populations between regions, for example through the development of a discrete spatial diffusion model.

The case study presented in Section \ref{sec:cyprus} demonstrates the utility of the framework to produce population estimates with a simple model and a single proxy, while ensuring that noise is not misinterpreted as signal---allowing for confidence in the validity of the results. The model extensions discussed above hold the potential to capture more complex phenomena with more complex datasets, which would allow for better model construction and comparison, and could enable the testing of demographic, social, and technological theories with archaeological data. Many questions will require additional data to answer, but an analysis process with clear uncertainty quantification and model assumptions can allow for further constraints based on existing data, can be used to guide the collection of new data that will be most informative for key questions, and can ensure that the maximum possible insight is gained from those new data.

\section*{Acknowledgements}
DL and FC's research contribution was supported by the European Research Council under the European Union's Horizon 2020 research and innovation program for the project `CLaSS - Climate, Landscape, Settlement and Society: Exploring Human Environment Interaction in the Ancient Near East' (grant number 802424, award holder: Dan Lawrence).

\printbibliography

\end{document}